%% file: extended.tex
\documentclass{llncs}

\input{preamble.tex}

\usepackage{makeidx}  
\begin{document}
\frontmatter          
\pagestyle{plain}
\mainmatter              
\title{Automatic Error Localization for Software using Deductive Verification
\thanks{This work was supported by the European Commission through project 
STANCE (31775) and the Austrian Science Fund (FWF) through project RiSE 
(S11406-N23).}}
\titlerunning{Automatic Error Localization for Software using Deductive
Verification}
%
\author{Robert K\"{o}nighofer \and
        Ronald Toegl \and
        Roderick Bloem}
%
%
%
\institute{IAIK, Graz University of Technology, Austria.}

\maketitle              

\begin{abstract}
Even competent programmers make mistakes.  Automatic verification can detect
errors, but leaves the frustrating task of finding the erroneous line of code to
the user.
This paper presents an automatic approach for identifying potential error
locations in software.  It is based on a deductive verification engine,
which detects errors in functions annotated with pre- and post-conditions. Using
an automatic theorem prover, our approach finds expressions in the code that can
be modified such that the program satisfies its specification.  Scalability is
achieved by analyzing each function in isolation.  We have implemented our
approach in the widely used \frama framework and present first experimental
results. This is an extended version of~\cite{conf}, featuring an additional
appendix.
\end{abstract}

\section{Introduction}

Formal verification attempts to detect mismatches between a program and its 
specification automatically.  However, the time-consuming work of locating and 
fixing detected bugs is usually performed manually. At the same time, the 
diagnostic information provided by the tools is often limited.  While model 
checkers commonly provide counterexamples, deductive software verification 
engines usually only give yes/no (or worse: only yes/maybe) answers. Analyzing 
a proof or witness given by the underlying theorem prover is usually not a 
viable option.

In this work, we strive to lessen this usability defect in the context of
deductive software verification~\cite{CuoqKKPSY12}.  This approach assumes that
source code is annotated with pre- and post-conditions.  It computes a set of
\emph{proof obligations}, i.e., formulas that need to be proven to attest
correctness.  These formulas are then discharged by an automatic theorem prover.
Scalability is achieved by analyzing functions in isolation.  We extend this
verification flow such that the tool does not only report the
existence of an error, but also pinpoints its location.

Our solution assumes that some code expression is faulty.  This fault model is
fine-grained and quite general.  If verification of a function fails, we iterate
over each expression in this function and analyze if it can be modified such
that the function satisfies its contract for all inputs.
If so, we report this expression as potential error
location. Expressions that cannot be modified such that the error
goes away do not have to be analyzed by the developer when trying to fix the
error. We have implemented a proof-of-concept in
\frama~\cite{CuoqKKPSY12}, and provide first experimental results comparing our 
approach to \forensic~\cite{BloemDFFHKRRS12} and
\bugassist~\cite{JoseM11}.

\textbf{Related work.}
Our fault model has been successfully applied 
before~\cite{KonighoferB11,BloemDFFHKRRS12}: This approach also checks 
repairability of expressions, but only for fixed inputs.  It uses assertions as 
specification, and SMT solvers as reasoning engines. \cite{GriesmayerSB07} is 
similar but uses a model checker. In~\cite{JoseM11} a MAX-SAT engine is used. 
Our work resolves many drawbacks of these existing works: pre- and 
post-conditions are more powerful than assertions, we check repairability for 
\emph{all} inputs, and we achieve scalability by analyzing functions in 
isolation. Model-based diagnosis~\cite{Reiter87} has already been applied in 
many settings (cf.~\cite{KonighoferB11}).  Our approach is similar (we also 
check repairability), but focuses on single-fault diagnoses to avoid floods of 
diagnoses. Dynamic methods~\cite{JonesH05} rely on the quality of available test 
cases.  In contrast, our method is purely formal.  This is an extended 
version of~\cite{conf}, featuring an additional appendix with more 
detailed experimental results.



\section{Automatic Error Localization}\label{sec:loc}

\subsection{Fault Models}\label{sec:fault}

Intuitively, a fault model defines what can go wrong in a program, thereby
inducing a set of candidate error locations. An error
localization algorithm can then
\begin{wrapfigure}[3]{r}{0.6\textwidth}
\vspace{-0.85cm} 
    \includegraphics[width=0.6\textwidth]{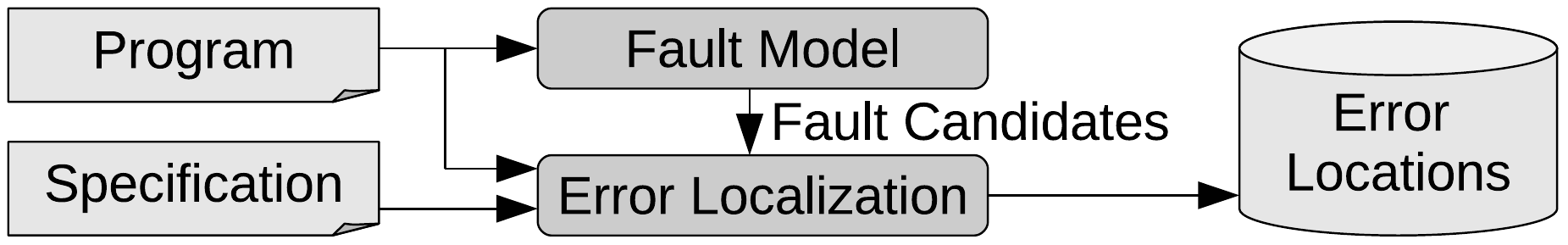}
\end{wrapfigure}
decide which of these candidates can actually be responsible for the detected
problem. A good fault model needs to balance conflicting objectives: it should
cover many errors, be fine-grained, allow for efficient error localization and
not yield too many spurious error locations. Existing approaches include fault
patterns~\cite{LarusBDDFPRV04} specifying common bugs, mutation-based fault
models~\cite{DebroyW10} assuming that the error is a small syntactic change,
and faulty expressions~\cite{GriesmayerSB07,KonighoferB11} assuming that the
control structure is correct but some code expression may be wrong.
In this work we use faulty expressions because this fault model is fine-grained,
more generic than mutation-based models, more automatic than fault
patterns, and still allows for efficient error localization, as shown below.

\subsection{Basic Idea for Error Localization}

Our approach is inspired by~\cite{GriesmayerSB07,KonighoferB11}:  An expression
in the source code is a potential error location if it can be replaced such that
the detected error is resolved.

\begin{exa} \label{ex:1}
The program on the right is supposed to compute the maximum of
\begin{wrapfigure}[5]{r}{0.38\textwidth}
\vspace{-0.90cm} 
\begin{lstlisting}
/*@ensures \result >= b;@*/
int max(int a, int b) {
  int r = a;               /*E\label{lst_ex_orig_2}E*/
  if(b > a)                /*E\label{lst_ex_orig_3}E*/
    r = a; //correct: r = b/*E\label{lst_ex_orig_4}E*/
  return r; }              /*E\label{lst_ex_orig_5}E*/
\end{lstlisting}
\end{wrapfigure}
\texttt{a} and \texttt{b}, but contains a
bug in line~\ref{lst_ex_orig_4}.
The post-condition \verb'\result >= b' is incomplete but
sufficient
to detect the bug: it is violated if $b > a$.
Our fault model (incorrect expressions) identifies 4 candidate error
locations:
Candidate $C_1$ is the expression ``\texttt{a}'' in line~\ref{lst_ex_orig_2},
$C_2$ is ``\texttt{b > a}'' in line~\ref{lst_ex_orig_3},
$C_3$ is the ``\texttt{a}'' in line~\ref{lst_ex_orig_4}, and
$C_4$ is the ``\texttt{r}'' in line~\ref{lst_ex_orig_5}.
Neither $C_1$ nor $C_2$ are error locations.
$C_1$ cannot be changed to satisfy the post-condition because
\texttt{r} is overwritten with the incorrect value ``\texttt{a}''
if $b > a$.  If we change only $C_2$,
\verb'\result' will always be ``\texttt{a}'', which is incorrect if $b > a$.
$C_3$ and $C_4$ are
possible error locations, because these expressions can be replaced by
``\texttt{b}'' to make the program satisfy its specification. \qed

\end{exa}
\subsection{Realization with Deductive Verification}

\noindent
We now discuss how to answer such repairability questions automatically.
From a high-level perspective, most formal verification tools compute a
correctness condition $\textsf{correct}(\overline{i})$ in some logic, where
$\overline{i}$ is the vector of input variables of the program. Next, a solver 
checks if $\forall \overline{i}\scope
\textsf{correct}(\overline{i})$ holds.  If not, an error has been detected.
Deductive verification tools like the \wpp plug-in of \frama~\cite{CuoqKKPSY12}
follow this pattern by defining $\textsf{correct}$ as implication: if the
pre-condition of a function holds, then the function must satisfy its
post-condition.  Loops are handled with user-provided invariants, and a
theorem prover checks $\forall \overline{i}\scope
\textsf{correct}(\overline{i})$.  In practice, \textsf{correct} may be composed
of parts that can be solved independently.

If a function is incorrect, we compute if a certain expression $C$ is a
potential error location as follows.  First, we replace $C$ by a placeholder $c$
for a new expression.  Next, we compute the correctness condition
$\textsf{correct}(\overline{i}, c)$, which depends now also on $c$. Finally, $C$
is a potential error location if $\forall \overline{i}: \exists c:
\textsf{correct}(\overline{i},c)$.  This formula asks if expression $C$ can, in
principle, be replaced such that the function satisfies its contract.  For every
input $\overline{i}$, there must exist a value $c$ to which the replacement of
$C$ evaluates such that the function behaves as specified.
Note that this approach can, in principle, also compute a repair if the
underlying theorem prover can produce a witness in form of a Skolem function
for the $c$ variable. However, this feature is not supported by our current
implementation.

\begin{exa}
We continue Example~\ref{ex:1}.  We check if expression $C_1$ is a potential
\begin{wrapfigure}[5]{r}{0.38\textwidth}
\begin{minipage}{0.38\textwidth}
\vspace{-0.90cm} 
\begin{lstlisting}
/*@ensures \result >= b;@*/
int max(int a, int b) {
  int r = /*E\mrk{c1}E*/;
  if(b > a)
    r = a; //correct: r = b
  return r; }
\end{lstlisting}
\end{minipage}
\end{wrapfigure}
error location by replacing it with
a placeholder $c_1$, as shown on the right.
Next, we compute
$\textsf{correct}(a, b, c_1) = (b \leq a) \wedge (c_1 \geq b)$
using deductive verification.
$C_1$ is not an error location because
$\forall
a,b\scope \exists c_1
\scope\textsf{correct}(a, b, c_1)$
is false.
When replacing $C_3$ we get $\textsf{correct}(a, b, c_3) = (b \leq a) \vee (c_3
\geq b)$.
We have
that $\forall a,b\scope \exists c_3\scope (b \leq a) \vee (c_3 \geq b)$,
so
$C_3$ is a potential error location --- as expected.
\qed

\end{exa}

\subsection{Implementation in \frama}\label{sec:impl}

We implemented our error localization approach as a proof of concept in the
\wpp plug-in of the widely used software verification framework
\frama~\cite{CuoqKKPSY12}. We discuss implementation challenges and reasons for
imperfect diagnostic resolution.\\
\noindent
\textbf{Instrumentation.}  \frama normalizes the source code while parsing it
into an Abstract Syntax Tree (AST). For instance, it decomposes complicated
statements using auxiliary variables.  Our instrumentation, replacing
candidate expressions by a placeholder $c$, operates on this
normalized AST. This makes it robust when
handling complicated constructions.  The
disadvantage is that our approach may report error locations that are only
present in the normalization.
However, we do not consider this a severe usability issue, because the line
number in the original code is available, and \frama presents the normalized
source code and how it links to the original source code in its GUI.\\
\noindent
\textbf{Computation of $\textsf{correct}(\overline{i},c)$.}
Internally, the \wpp plug-in of \frama performs simplifications that may rewrite or eliminate our
newly introduced placeholder $c$, and thus, we cannot use  \wpp a black-box to
compute the correctness formula $\textsf{correct}(\overline{i},c)$ after
instrumentation. 
We solve this issue by extending \frama's memory model such that the placeholder
$c$ is not touched by simplifications.\\
\noindent
\textbf{Quantification.}
Once we have $\textsf{correct}(\overline{i}, c)$, we need to add the quantifier
prefix $\forall \overline{i}\scope \exists c$.  Unfortunately,
$\textsf{correct}$ may also contain auxiliary variables $\overline{t}$ that
express values of variables at specific program points.  Intuitively,
$c$ should not depend on variables that are assigned later in the program.  This
would violate the causality and lead to false-positives. Hence,
we need to separate the variables of $\textsf{correct}$ to construct the
formula
$\forall \overline{i}\scope \exists c\scope \forall \overline{t}\scope
\textsf{correct}(\overline{i}, \overline{t}, c)$.
This is done by computing the input variables (parameters and
globals) of the function under analysis
and linking them to the corresponding variables in the formula.\\
\noindent
\textbf{Axiomatization.}
\wpp uses axiomatized functions and predicates in
\textsf{correct}.  For instance, for $a < b$ it writes
$\text{zlt}(a,b)$, where the predicate $\text{zlt}: \mathbb{Z} \times \mathbb{Z}
\rightarrow \mathbb{B}$ is axiomatized as
$\forall x,y: (zlt(x,y) \rightarrow x<y) \wedge
               (\neg zlt(x,y)\rightarrow x\geq y)$.
In our experiments we observed cases where the automatic theorem prover
(\altergo) could not decide formulas when using the
axiomatization, but had no difficulty when the axiomatized predicates and
functions are replaced by the corresponding
native operators.  Hence, we modified the interface to the theorem prover such
that formulas do not contain axiomatized functions and predicates, where
possible.\\
\noindent
\textbf{Diagnostic Resolution.}
Our implementation is neither guaranteed to be sound (it may produce spurious
error locations) nor complete (it may miss potential error locations).  The
reasons are:
\begin{itemize}
\item The theorem prover may time-out or return ``Unknown'' if it could neither
prove nor disprove the formula.  We treat such verdicts as if the program was
incorrect (a choice justified by experience), which results in
incompleteness.
\item Instead of one monolithic formula \textsf{correct}, \wpp may compute
multiple formulas that are checked independently.  In error localization, we
also check each formula in isolation.  This is weaker than
checking the conjunction, i.e., can result in spurious error
locations, but increases efficiency.
\item Incomplete specifications can result in spurious error locations.
\item The bug may not match our fault model. E.g., code may be
missing or the control flow may be incorrect.  This results
in missed error locations.
\end{itemize}

\section{First Experimental Results}\label{sec:exp}

Despite the potential imprecisions discussed in the last section, our 
implementation produces meaningful results. We evaluated our proof-of-concept 
implementation\footnote{See 
\url{www.iaik.tugraz.at/content/research/design_verification/others/}.} on the 
widely used \tcas benchmark~\cite{Siemens},
which implements an aircraft traffic collision
avoidance system in 180 lines of C code.  It comes in 41
faulty versions that model realistic bugs.  We annotated all functions with
contracts.

\subsection{Performance Evaluation}

We compare the execution time and effectiveness of our approach with that of
\forensic~\cite{KonighoferB11,BloemDFFHKRRS12} and \bugassist~\cite{JoseM11} on
an ordinary laptop.\footnote{Table~1 in the Appendix gives
more details to our performance results.} For our new approach, the error
localization time (at most 129 [s], 37 [s] on average) is acceptable for all
\tcas instances. For 37\% of the cases, the execution time increases by only
$<$40\% when going from error detection to localization. \forensic is slightly
faster on average (17 [s]) but the median runtime is on par (16 vs. 18 [s]).
With 7 [s] on average, \bugassist is even faster. Although only 66\% of
the benchmarks match our fault model, errors were successfully located in
90.2\%. While \forensic and \bugassist reported 15 error locations on average,
our approach reported only 3.5.  Thus, in our experiments, our tool provides
much higher accuracy with only slightly longer runtime. The user has to examine
only a few expressions in the code, which can speed-up debugging significantly.

\subsection{Examples}

This section investigates the reported error locations for a few \tcas versions.

\noindent
\textbf{Version 7.}
A constant is
changed from $500$ to $550$ in an initialization function.
Our tool reports exactly this constant \texttt{550} as the only possible error
location.
This takes 6 seconds, whereof 5.1 seconds are spent on error detection.

\noindent
\textbf{Version 9.}
This version contains the following function:
\begin{wrapfigure}[6]{r}{0.76\textwidth}
\begin{minipage}{0.76\textwidth}
\vspace{-0.95cm} 
\begin{lstlisting}[firstnumber=119]
bool NonCrossBiasedDescend() {
  bool r;
  if (InhibitBiasedClimb() >= DwnSep) { /*E\label{lst_tcas9_bug}E*/
    r = OwnBlTh() && VerSep >= MSEP && DwnSep >= ALIM();/*E\label{lst_tcas9_res1}E*/
  } else {
    r = !(OwnAbTh()) || (OwnAbTh() && UpSep >= ALIM());/*E\label{lst_tcas9_res2}E*/
  }
  return r; }
\end{lstlisting}
\end{minipage}
\end{wrapfigure}
The correct program has a ``\texttt{>}'' instead of the ``\texttt{>=}'' in
line \ref{lst_tcas9_bug}. Our tool
reports two potential error locations:
\texttt{tmp\_6 >= DwnSep} in line \ref{lst_tcas9_bug}, and \texttt{tmp\_1} in
line \ref{lst_tcas9_res1}.
This output looks cryptic because the code
has been normalized by \frama. \texttt{tmp\_6} is an auxiliary variable that
stands for \texttt{InhibitBiasedClimb()}. This is shown in the GUI.  Hence,
the first error location is just what we expect.  \texttt{tmp\_1} holds the
value for \texttt{r} in line~\ref{lst_tcas9_res1}. This value can be changed to
satisfy the specification for all inputs as well.  Hence, it is also reported.
\texttt{NonCrossBiasedDescend()} is not long, but contains complex logic.
Analyzing this logic to locate a bug can be cumbersome.  The diagnostic
information provided by our approach helps.

\noindent
\textbf{Version 14} changes \texttt{MAXDIFF} (a preprocessor macro) from
\texttt{600} to \texttt{600+50}.
Our tool reports two possible error locations: \texttt{VerSep >
600+50} in line \ref{lst_tcas14_bug} and \texttt{OtherCap == 1} in line
\ref{lst_tcas14_rep1} of function \texttt{altSepTest}, which is shown below.
The first one
pinpoints exactly the problem. Note that \texttt{altSepTest()} is all but trivial.
\begin{wrapfigure}[11]{r}{0.72\textwidth}
\begin{minipage}{0.72\textwidth}
\begin{lstlisting}[firstnumber=165]
int altSepTest() {
  bool en, eq, intentNotKnown, needUpRA, needDwnRA;
  en = HConf && OwnTrAlt <= OLEV && VerSep > MAXDIFF;/*E\label{lst_tcas14_bug}E*/
  eq = OtherCap == TCAS_TA;                          /*E\label{lst_tcas14_rep1}E*/
  intentNotKnown = TwoRepValid && OtherRAC == NO_INT;
  int altSep = UNRESOLVED;
  if (en && ((eq && intentNotKnown) || !eq)) {
    needUpRA = NonCrossBiasedClimb() && OwnBlTh();
    needDwnRA = NonCrossBiasedDescend() && OwnAbTh();
    if(needUpRA && needDwnRA) altSep = UNRESOLVED;
    else if (needUpRA)        altSep = UPWARD_RA;
    else if (needDwnRA)       altSep = DOWNWARD_RA;
    else                      altSep = UNRESOLVED;
  }
  return altSep; }
\end{lstlisting}
\end{minipage}
\end{wrapfigure}
If verification fails, tracking
 down this bug can be a very
time-consuming and frustrating task. By checking only the reported locations, we
can significantly reduce the manual work to fix the bug.
Thus, the reported error locations are usually both meaningful and helpful.

\section{Conclusions}\label{sec:conc}

Tracking down a subtle program error in large source code is --- like finding a
needle in a haystack --- a tedious task.  We have extended a widely used
deductive software verification engine so that it can report expressions that
may be responsible for incorrectness.  We evaluated our proof-of-concept
implementation on a few examples and conclude that our approach is viable and
gives fast and clear guidance to developers on the location of program 
defects.\\

\noindent
\textbf{Acknowledgment.}
We thank Lo\"{\i}c Correnson and the \frama team
for their support with our proof-of-concept implementation.

\bibliography{refs}

\newpage
\appendix
\section*{Appendix}
Table~\ref{table:evaluation} gives more details to our performance results on
the \tcas benchmarks. Column~1 indicates if the error in this version of the
benchmark matches our fault model. Even if this is not the case, our approach
can often compute meaningful error locations.  Column~2 lists the execution time
for error detection. Column~3 gives the time for error localization including
error detection. Column~4 gives the number of candidate expressions identified
by our fault model. The number of potential error locations that have been
reported by our implementation is listed in Column~5. The Columns 6 and 7 show
the execution time for error localization (including error detection) and the
number of reported (potential) error locations for the approach
of~\cite{KonighoferB11}, which is implemented in the tool
\forensic~\cite{BloemDFFHKRRS12}.  This approach can be run in two modes: the
conservative mode may miss error locations, the non-conservative mode may find
spurious locations.  We used the non-conservative mode because otherwise no
error locations are found for several benchmark versions.  Furthermore, we let
\forensic compute single-fault diagnoses only. Otherwise, the number of
diagnoses grows to several hundreds for certain benchmark versions.  The last
two columns show the same information for the approach of~\cite{KonighoferB11},
which has been implemented in the tool \bugassist. All experiments were
performed on a notebook with an Intel Core i5-3320M processor running at 2.6
GHz, 8GB of RAM, and a 64 bit Linux operating system.  The memory consumption
was insignificant in our experiments.

\begin{table}[ht]
\setlength{\tabcolsep}{3.7pt}
\renewcommand{\arraystretch}{0.90}
\begin{center}
\caption{Detailed performance results.}
\label{table:evaluation}
\begin{tabular}{ccccccccccccc}
\toprule[1.3pt]
Column    &1 &&2 &3 &4 &5 &&6 &7 &&8 &9\\
\midrule
\multicolumn{2}{c}{\tcas Benchmark}
&&\multicolumn{4}{c}{Our new approach}
&&\multicolumn{2}{c}{\forensic~\cite{KonighoferB11}}
&&\multicolumn{2}{c}{\bugassist~\cite{JoseM11}}\\
\cmidrule{1-2} \cmidrule{4-7} \cmidrule{9-10} \cmidrule{12-13}
          &Matches
          &&Error
          &Error
          &Nr. of
          &Nr. of
          &&Error
          &Nr. of
          &&Error
          &Nr. of
          \\
          Version
          &Fault
          &&Det.
          &Loc.
          &Cand-
          &Loc.
          &&Loc.
          &Loc.
          &&Loc.
          &Loc.
          \\
          &Model
          &&Time
          &Time
          &idates
          &Rep.
          &&Time
          &Rep.
          &&Time
          &Rep.
          \\
   &[-]  &&[s]&[s]&[-]&[-]    &&[s] &[-]   &&[s] &[-]\\
\cmidrule{1-2} \cmidrule{4-7} \cmidrule{9-10} \cmidrule{11-13}
1  & Yes &&  5.2&   20& 10&  4&&   19& 22  && 7.8& 16\\
2  & Yes &&  5.2&  6.9&  4&  2&&   14& 15  && 9.8& 17\\
3  &  No &&  4.5&  111& 35& 13&&   15& 12  &&  10& 17\\
4  &  No &&  5.3&   22& 10&  0&&   17& 20  && 7.5& 17\\
5  &  No &&  4.3&   84& 33&  7&&   27& 12  && 4.3& 18\\
6  & Yes &&  5.2&  6.0&  2&  2&&   17& 20  && 5.6& 17\\
7  & Yes &&  5.1&  6.0&  4&  1&&   15& 11  && 7.6& 17\\
8  & Yes &&  5.2&  6.8&  4&  1&&   15& 12  && 8.1& 15\\
9  & Yes &&  5.2&   20& 10&  2&&   14& 19  && 8.6& 13\\
10 & Yes &&  5.3&  6.9&  4&  4&&   23& 18  &&  11& 18\\
11 & Yes &&  5.3&  6.9&  4&  4&&   32& 13  && 6.3&  9\\
12 &  No &&  4.4&   89& 35&  5&&   28& 21  && 5.5& 18\\
13 & Yes &&  4.3&  111& 35&  8&&   20& 12  && 6.7& 16\\
14 & Yes &&  4.7&  104& 35&  2&&   15&  3  && 7.0&  8\\
15 & Yes &&  5.3&   35& 20&  6&&   18& 11  && 4.4& 18\\
16 & Yes &&  5.2&  6.8&  4&  1&&   15& 11  && 7.7& 16\\
17 & Yes &&  5.2&  6.8&  4&  1&&   14& 11  && 8.0& 16\\
18 & Yes &&  5.2&  6.8&  4&  1&&   14& 11  && 7.8& 16\\
19 & Yes &&  5.2&  7.0&  4&  1&&   15& 11  && 7.9& 16\\
20 & Yes &&  5.3&   20& 10&  2&&   14& 21  && 7.3& 17\\
21 & Yes &&  5.0& 18.4& 10&  2&&   11& 21  &&  10& 17\\
22 & Yes &&  5.1& 18.3& 10&  2&&   11& 18  && 7.1& 16\\
23 & Yes &&  5.2& 18.4& 10&  2&&   11& 19  && 8.7& 13\\
24 & Yes &&  5.1& 18.5& 10&  2&&   11& 21  &&  10& 17\\
25 & Yes &&  5.2&   20& 10&  3&&   18& 21  && 6.8& 16\\
26 &  No &&  4.2&   93& 33&  8&&   20& 12  && 6.2& 17\\
27 &  No &&  4.3&   87& 33&  7&&   26& 12  && 4.2& 18\\
28 & Yes &&  5.3&   10&  4&  2&&   20& 10  && 8.2& 14\\
29 & Yes &&  5.2&  5.6&  1&  1&&   14& 14  && 7.4& 13\\
30 & Yes &&  5.5&  6.0&  2&  2&&   14& 14  && 9.1& 17\\
31 &  No &&  4.9&  129& 48& 10&&   16& 15  && 3.4& 15\\
32 &  No &&  4.8&  102& 48&  7&&   16& 15  && 3.2& 17\\
33 &  No &&  5.1&  6.8&  4&  0&&   26& 12  && 0.3&  1\\
34 &  No &&  4.2&   30& 35&  1&&   28& 11  && 4.3& 18\\
35 & Yes &&  4.4&  7.0&  4&  2&&   18& 10  &&  10& 18\\
36 & Yes &&  5.0&  129& 35& 13&&   19& 22  && 4.3& 15\\
37 &  No &&  5.2&  6.1&  2&  0&&   19& 12  && 7.4& 15\\
38 &  No &&  5.2&  5.2&  0&  0&&    2&  0  && 0.3&  1\\
39 & Yes &&  5.2&   20& 10&  3&&   17& 21  && 7.0& 16\\
40 &  No &&  4.2&   80& 41&  8&&   18& 19  && 7.6& 16\\
41 &  No &&  5.0&   17&  9&  3&&   16& 19  && 6.4& 16\\
\cmidrule{1-2} \cmidrule{4-7} \cmidrule{9-10} \cmidrule{12-13}
average&66\,\%&&5.0&37& 15&3.5&&   17& 15  && 6.9& 15\\
median&       &&5.2&18& 10&  2&&   16& 14  && 7.4& 16\\
\bottomrule[1.3pt]
\end{tabular}
\end{center}
\end{table}

\end{document}

%% file: preamble.tex
\usepackage{amsthm}
\usepackage{booktabs}
\usepackage{rotating}
\usepackage{color}
\usepackage{caption}
\usepackage{listings}
\usepackage{xspace}
\usepackage{hyperref}
\usepackage{amsfonts}
\usepackage{amsmath}

\usepackage{amssymb}
\usepackage{hyperref}
\usepackage{stmaryrd}
\usepackage{enumerate}
\usepackage{paralist} 
\usepackage{wrapfig} 

\newcounter{exacounter}
\newenvironment{exa}{
\refstepcounter{exacounter}
\smallskip\noindent
\textbf{Example \theexacounter.}
}{}

\definecolor{darkgreen}{rgb}{0,0.5,0}
\definecolor{darkblue}{rgb}{0,0,.5}
\definecolor{lightblue}{rgb}{0.8,0.85,1}
\definecolor{mygray}{gray}{.3}
\lstdefinelanguage{my_c}{morekeywords={
                  if, else, while, do, for, continue, break, return, int},
                           keywordstyle=\bf\color{darkblue},
                           directivestyle=\color{darkblue},
                           commentstyle=\color{darkgreen},
                           sensitive=true,
                           morecomment=[l]{//},
                           morecomment=[s]{/*}{*/},
                           morestring=[b]',
                           morestring=[b]",
                           moredelim=*[l][\color{darkgreen}]\#
                          }[keywords,comments,strings,directives]
\newcommand{\mrk}[1]{\textbf{\color{red}#1}}

\newcommand{\altergo}{\textsf{AltErgo}\xspace}
\newcommand{\frama}{\textsf{Frama-C}\xspace}
\newcommand{\forensic}{\textsf{FoREnSiC}\xspace}
\newcommand{\bugassist}{\textsf{Bug-Assist}\xspace}
\newcommand{\wpp}{\textsf{WP}\xspace}
\newcommand{\tcas}{\textsf{TCAS}\xspace}
\DeclareMathOperator{\scope}{\mathbin{:}}

%% file: extended.bbl
\begin{thebibliography}{10}

\bibitem{BloemDFFHKRRS12}
R.~Bloem, R.~Drechsler, G.~Fey, A.~Finder, G.~Hofferek, R.~K{\"o}nighofer,
  J.~Raik, U.~Repinski, and Andr{\'e} S{\"u}lflow.
\newblock {FoREnSiC} - {A}n automatic debugging environment for {C} programs.
\newblock In {\em HVC'12}. Springer, 2012.

\bibitem{CuoqKKPSY12}
P.~Cuoq, F.~Kirchner, N.~Kosmatov, V.~Prevosto, J.~Signoles, and B.~Yakobowski.
\newblock {Frama-C} - {A} software analysis perspective.
\newblock In {\em SEFM'12}. Springer, 2012.

\bibitem{DebroyW10}
V.~Debroy and W.~E. Wong.
\newblock Using mutation to automatically suggest fixes for faulty programs.
\newblock In {\em ICST'10}. IEEE, 2010.

\bibitem{GriesmayerSB07}
A.~Griesmayer, S.~Staber, and R.~Bloem.
\newblock Automated fault localization for {C} programs.
\newblock {\em Electr. Notes Theor. Comput. Sci.}, 174(4):95--111, 2007.

\bibitem{JonesH05}
J.~A. Jones and M.~J. Harrold.
\newblock Empirical evaluation of the tarantula automatic fault-localization
  technique.
\newblock In {\em ASE'05}. ACM, 2005.

\bibitem{JoseM11}
M.~Jose and R.~Majumdar.
\newblock Cause clue clauses: error localization using maximum satisfiability.
\newblock In {\em PLDI'11}, pages 437--446. ACM, 2011.

\bibitem{KonighoferB11}
R.~K{\"o}nighofer and R.~Bloem.
\newblock Automated error localization and correction for imperative programs.
\newblock In {\em FMCAD'11}. IEEE, 2011.

\bibitem{conf}
R.~K{\"o}nighofer, R.~Toegl, and R.~Bloem.
\newblock Automatic error localization for software using deductive
  verification.
\newblock In {\em HVC'14}. Springer, 2014.
\newblock To appear.

\bibitem{LarusBDDFPRV04}
J.~R. Larus, T.~Ball, M.~Das, R.~DeLine, M.~F{\"a}hndrich, J.~D. Pincus, S.~K.
  Rajamani, and R.~Venkatapathy.
\newblock Righting software.
\newblock {\em IEEE Softw.}, 21(3):92--100, 2004.

\bibitem{Reiter87}
R.~Reiter.
\newblock A theory of diagnosis from first principles.
\newblock {\em Art. Int.}, 32(1):57--95, 1987.

\bibitem{Siemens}
{Siemens benchmark suite}.
\newblock \url{pleuma.cc.gatech.edu/aristotle/Tools/subjects}.

\end{thebibliography}
